\input harvmac
\input epsf
\batchmode
  \font\bbbfont=msbm10
\errorstopmode
\newif\ifamsf\amsftrue
\ifx\bbbfont\nullfont
  \amsffalse
\fi
\ifamsf
\def\IR{\hbox{\bbbfont R}}
\def\IZ{\hbox{\bbbfont Z}}
\def\IF{\hbox{\bbbfont F}}
\def\IP{\hbox{\bbbfont P}}
\else
\def\IR{\relax{\rm I\kern-.18em R}}
\def\IZ{\relax\ifmmode\hbox{Z\kern-.4em Z}\else{Z\kern-.4em Z}\fi}
\def\IF{\relax{\rm I\kern-.18em F}}
\def\IP{\relax{\rm I\kern-.18em P}}
\fi
\def\np#1#2#3{Nucl. Phys. B {#1} (#2) #3}
\def\pl#1#2#3{Phys. Lett. B {#1} (#2) #3}

\def\ev#1{\langle#1\rangle}

\def\O{{\cal O}}
\def\Ot{{\cal O}_{\tau}}
\def\Otbar{{\cal O}_{\overline \tau}} 
\def\f#1#2{\textstyle{#1\over #2}}
\def\lfm#1{\medskip\noindent\item{#1}}
\lref\review{See O. Aharony,
S.S. Gubser, J. Maldacena, H. Ooguri, and Y. Oz, hep-th/9905111.}
\lref\malda{J.M. Maldacena, hep-th/9711200.}
\lref\GKP{S.S. Gubser, I. Klebanov, and A.M. Polyakov, hep-th/9802109,
\pl{428}{1998}{10}.}
\lref\EW{E. Witten, hep-th/9802150, Adv. Theor. Math. Phys. 2 (1998)
253.}
\lref\kleba{I. Klebanov, hep-th/9702076, \np{496}{1987}{231}.}
\lref\GM{M. Gunaydin and N. Marcus, Class. Quant. Grav. 2, L11.}
\lref\GHKK{S.S. Gubser, A. Hashimoto, I.R. Klebanov, and M. Krasnitz,
hep-th/9803023, \np{526}{1998}{393}.}
\lref\SGAH{S.S. Gubser and A. Hashimoto, hep-th/9805140.}
\lref\HorStrom{G. Horowitz and A. Strominger, \np{B360}{1991}{197}.}
\lref\minwalla{S. Minwalla, hep-th/9803053, JHEP 9810 (1998) 002.}
\lref\malda{J.M. Maldacena, hep-th/9711200, Adv. Theor. Math. Phys. 2
(1998) 231.}
\lref\GKP{S.S. Gubser, I. Klebanov, and A.M. Polyakov, hep-th/9802109,
\pl{428}{1998}{105}.}
\lref\EW{E. Witten, hep-th/9802150, Adv. Theor. Math. Phys. 2 
(1998) 253.}
\lref\kleba{I. Klebanov, hep-th/9702076, \np{496}{1987}{231}.}

\lref\SW{L. Susskind and E. Witten, hep-th/9805114.}
\lref\sixteen{N. Seiberg, hep-th/9705117, Nucl. Phys. Proc. Suppl. 67
(1998) 158.}
\lref\KI{K. Intriligator, hep-th/9811047, \np{551}{1999}{575}.}
\lref\WS{W. Skiba, hep-th/9907088.}
\lref\NSback{N. Seiberg, hep-ph/9309335, \pl{318}{1993}{469}.}
\lref\GKT{S.S. Gubser, I.R. Klebanov, A.A. Tseytlin, hep-th/9703040,
\np{499}{1997}{217}.}
\lref\SGIK{S.S. Gubser, I.R. Klebanov, hep-th/9708005,
\pl{413}{1997}{41}.} 
\lref\oscvid{M. Gunaydin, P. van Nieuenhuizen, and N.P. Warner,
\np{255}{1985}{63}.} 
\lref\osciiid{M. Gunaydin and N.P. Warner, \np{272}{1986}{99}.}
\lref\IKEW{I.R. Klebanov and E. Witten, hep-th/9905104.}
\lref\MVS{S. Minwalla, M. Van Raamsdonk, N. Seiberg, hep-th/9912072.}
\Title{hep-th/9909082, UCSD/PTH-98/39, IASSNS-HEP-98/98}
{\vbox{\centerline{Maximally Supersymmetric RG Flows and AdS Duality
}}}
\medskip
\centerline{Kenneth Intriligator}
\vglue .5cm
\centerline{UCSD Physics Department}
\centerline{9500 Gilman Drive}
\centerline{La Jolla, CA 92093}

\bigskip
\noindent
We discuss four dimensional renormalization group flows which preserve
sixteen supersymmetries.  In the infra-red, these can be viewed as
deformations of the ${\cal N}=4$ superconformal fixed points by
special, irrelevant operators.  It is argued that the gauge coupling
beta function continues to vanish identically, for all coupling
constants and energy scales, for such RG flows.  In addition, the
dimensions of all operators in short supersymmetry representations are
constant along such flows.  This is compatible with a conjectured
generalization of the AdS/CFT correspondence which describes such
flows, e.g. the D3 brane vacuum before taking the near-horizon limit.
RG flows in three and six dimensions, preserving 16 supersymmetries,
are also briefly discussed, including a discussion of generalized
AdS/CFT duality for the M2 and M5 brane cases.  Finally, we discuss
maximally supersymmetric RG flows associated with non-commutative
geometry.

%\draftmode
\Date{9/99}           

\newsec{Introduction}
Four dimensional ${\cal N}=4$ Super-Yang-Mills theories have 32
conserved supercharges: 16 ordinary ones, $Q_\alpha ^I$ and $\overline
Q _{\dot \alpha I}$, and 16 additional superconformal supercharges,
$S_{\alpha I}$ and $\overline S^I_{\dot
\alpha}$, with $\alpha$, $\dot \alpha =1,2$ spinor indices and
$I=1\dots 4$ in the fundamental of the global $SU(4)_R$ symmetry. The
16 additional, superconformal supercharges are associated with the
fact that the theory is conformally invariant, with $\beta _\tau
(\tau)\equiv 0$ for arbitrary gauge coupling and theta angle $\tau
\equiv{\theta \over 2\pi}+4\pi i g_{YM}^{-2}$.  We will here be
interested in theories which preserve the 16 ordinary supersymmetries
but are not conformally invariant.  These are the maximally
supersymmetric renormalization group flows (introducing more
supersymmetries either makes the theory conformally invariant or
necessitates adding gravity and other higher spin fields).  

The infrared endpoints of such RG flows are the usual ${\cal N}=4$
superconformal theories with 32 supercharges.  The RG flows can be
viewed as these RG fixed points with additional perturbing operators,
which preserve the 16 supersymmetries and become irrelevant in the
infrared.  The least irrelevant such operator is a dimension 8
operator of the form $\Tr F^4 +\dots$ (the $\dots$ are terms related
by the 16 supersymmetries; in terms of ${\cal N}=1$ supersymmetry, it
is $\int d^4 \theta \Tr [W_\alpha ^2
\overline W_{\dot \alpha} ^2]+\dots$); this operator is 
an $SU(4)_R$ singlet.  There is another dimension 8, $SU(4)_R$
singlet, scalar operator, which also preserves 16 supersymmetries,
given by $\Tr F^2 \Tr F^2 +\dots$.  Because these two operators have
the same quantum numbers, it is exceedingly difficult to tell them
apart (see
\WS)-- so we will not bother doing so.  We will refer to either
operator, or a general mixture of them, as $\O _H$.

Other operators which preserve the 16 supersymmetries are of dimension
$8+n$, of the form $\Tr F^4\phi ^n+\dots $, and in the $SU(4)_R$
representation with Dynkin indices $(0,n,0)$, for arbitrary integer
$n\geq 0$.  Again, we will not bother distinguishing between these and
multi-trace analogs of these operators with the same quantum numbers.
All of these operators are in short supersymmetry representations and
are of the form $Q^4 \overline Q^4 O_{short}$.  There are also 16
supersymmetry preserving operators in long representations of the
supersymmetry algebra, whose dimensions can vary with $g_{YM}$; these
operators are of the form $Q^8 \overline Q^8 O_{long}$ and thus have
dimension larger than 8.

We will argue in the next section that certain properties known to
hold for the 4d ${\cal N}=4$ superconformal fixed points continue to
hold along RG flows which preserve the 16 supersymmetries.  In
particular, the gauge coupling $g_{YM}$ and theta angle do not change
with the energy scale $\mu$ along such flows, i.e. $\beta _\tau (\tau
) \equiv \mu {d\over d\mu}\tau =0$ for all $\tau$.  Even though the
theory is not scale invariant, there are exactly marginal operators,
$\Ot$ and $\Otbar$, which deform $\tau$ and $\overline
\tau$ respectively.  This is depicted in figure 1. More generally,
there are still short representations of the supersymmetry algebra and
the dimensions of all operators in such representations are
independent of the energy scale along these RG flows.  Briefly, the
argument is that the dimension of the stress tensor is not
renormalized, because it is a conserved current, and that this, along
with supersymmetry, is enough to fix all operator dimensions in all
short multiplets to also not be renormalized.

\midinsert
\epsfxsize=3in \epsfbox{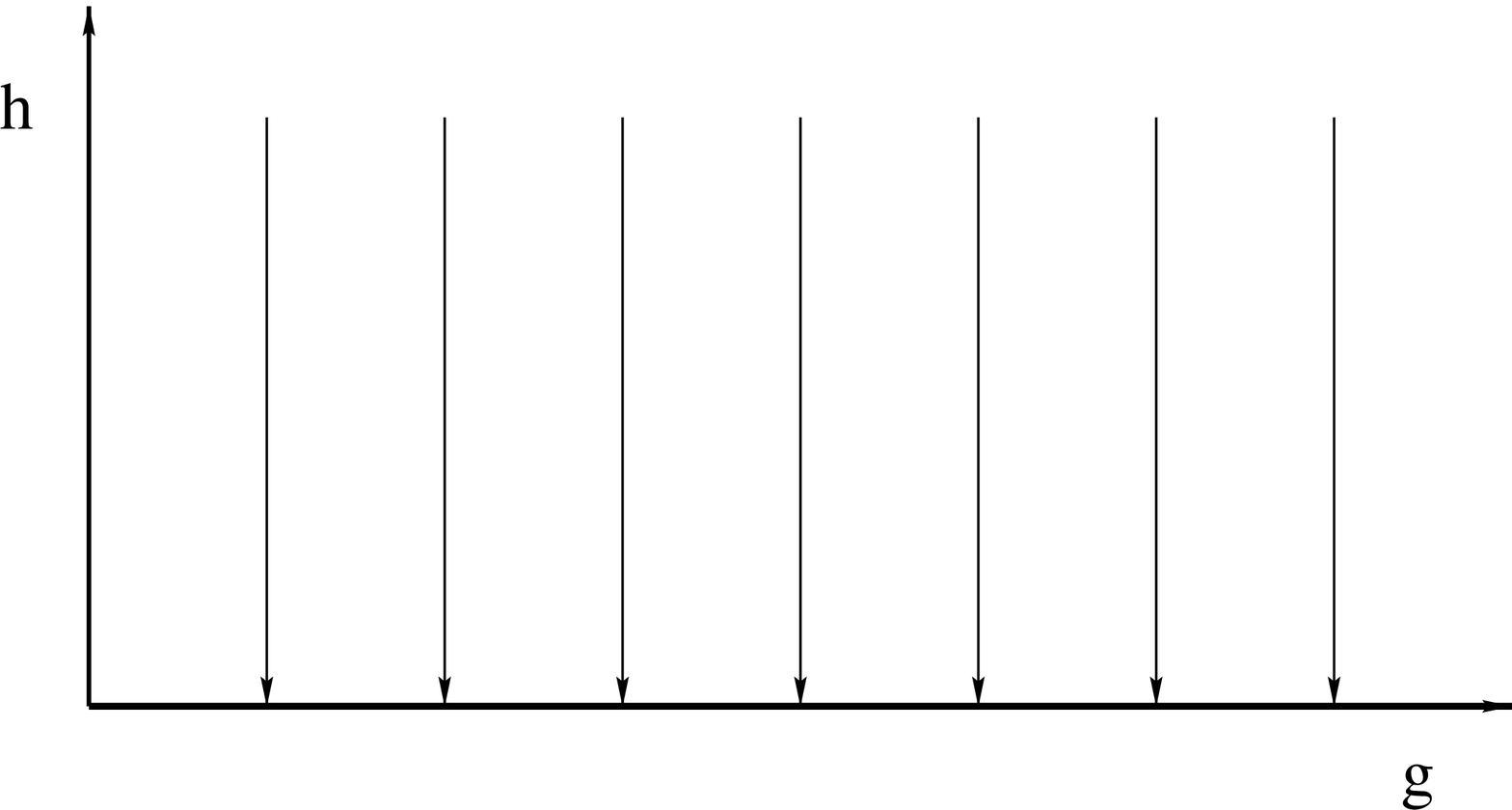} 

\noindent {\bf Figure 1.} RG flows to the IR, 
preserving 16 supersymmetries.  Coefficients $h$ of perturbing
operators flow to zero, while the gauge coupling $g$ remains constant
along the flow.
\endinsert

To give a concrete example of a RG flow preserving 16 supersymmetries,
consider deforming a ${\cal N}=4$ superconformal theory by moving away
{}from the origin of the moduli space of vacua, giving an expectation
value to the adjoint scalars $\phi$, which Higgses the gauge group $G$
to a subgroup such as $H\times U(1)$.  This deformation leads to a
non-trivial renormalization group flow, with 16 supersymmetries
preserved along the entire flow; the additional 16 superconformal
symmetries emerge in the UV and IR limits of the flow.  The UV limit
is the ${\cal N}=4$ superconformal theory with gauge group $G$ and the
IR limit is the ${\cal N}=4$ superconformal theory with gauge group
$H\times U(1)$.  The fact that the gauge coupling $\tau$ is exactly
(i.e. including all quantum effects, both perturbative and
non-perturbative) constant along such renormalization group flows is
well known: e.g. the Seiberg-Witten curve for the ${\cal N}=4$ theory
gives a gauge coupling which does not depend on the coordinates of the
Coulomb branch.  (This is a special case of the fact that the gauge
coupling does not depend on the Higgs branch moduli in ${\cal N}=2$
supersymmetric theories because of the decoupling there of vector and
hypermultiplets.)

Viewed from the IR, the above RG flow is the $H\times U(1)$ gauge
theory, perturbed by various irrelevant operators which preserve the
16 supersymmetries, such as the dimension 8 operator $O_H$ mentioned
above.  We will consider more general RG flows preserving 16
supersymmetries, without worrying about the UV starting point.  If
necessary, we can imagine that there is a UV cutoff, which preserves
the 16 supersymmetries, below the scale where the irrelevant operators
might blow up.

As will be discussed in sect. 3, D-branes and $AdS$ duality
\review\  motivate considering RG flows preserving 
16 supersymmetries.  In this section we will make contact with the
works \refs{\GHKK, \SGAH}.  We will discuss a conjectured extension of
the duality between field theories and gravity theories away from the
$AdS$ limit.  For example, IIB string theory in asymptotically flat
10d spacetime (with $N$ D3 branes infinitely far away), is conjectured
to be holographically dual to the UV limit of ${\cal N}=4$
supersymmetric $SU(N)$ gauge theory with a particular choice of
irrelevant operators.  The value of $N$ is arbitrary.

Section 4 is devoted to 3d RG flows with 16 preserved supercharges.
The IR fixed points are the 3d ${\cal N}=8$ superconformal theories
with 32 supercharges.  Several peculiarities are noted.  The flows
which are associated with $M$ theory vacua containing M2 branes are
discussed.

Section 5 is devoted to 6d RG flows with 16 preserved supercharges.  
The IR fixed points are the 6d ${\cal N}=(2,0)$ superconformal field
theories with 32 supercharges.  The flows which are associated with
$M$ theory vacua containing M5 branes are discussed. 

Section 6 discusses maximally supersymmetric RG flows associated with
non-commutative geometry.  The RG flow of fig. 1, with $g_{YM}$
constant, is verified in the string vacua which are proposed to be
holographically dual to the field theory RG flows.

In the appendices we list, for convenience, results obtained in 
\refs{\osciiid, \oscvid} (see also \minwalla) 
on the short representations of the superconformal group, with 32
supercharges, in 3d and 6d.

\newsec{The power of 16 supersymmetries}

Recall that the superconformal theory, with 32 supercharges, has two
types of operator representations: short and long.  The
representations are generated by a primary operator $\O _P$, along
with descendent operators, related to $\O _P$ by supersymmetry, of the
form $Q ^n\overline Q ^m \O _P$, and their conformal descendents.  It
should be understood that $Q^n\overline Q^m \O _P$ is a shorthand for
a nested graded commutator of the operator $\O _P$ with $n$ of the
supercharges $Q^I_\alpha$ and $m$ of the supercharges $\overline Q
_{I\dot \alpha}$, e.g. $Q^2\overline Q \O _P$ should be understood as
a shorthand for $[Q, \{Q,[\overline Q,\O _P]\}]$.  The remaining 16
superconformal supercharges, $S$ and $\overline S$, act on these
representations as lowering operators.  For the generic, long,
representation, the operators $Q^n\overline Q^m \O _P$ truncate at
$n\leq 8$ and $m\leq 8$ by Fermi statistics.  Taking $n>8$ or $m>8$ in
$Q ^n\overline Q ^m \O_P$ gives zero, up to a total derivative.  The
dimensions of long operators are not constrained by supersymmetry and
depend on $g_{YM}$ as well as the gauge group.

The short representations have the defining property that they instead
truncate at $n\leq 4$ and $m\leq 4$; they are the ${\cal N}=4$ version
of chiral superfields.  The spectrum of short representations was found
in \GM\ and their table, with our present field-theory operator
notation, can be found reproduced in
\KI.  The primary operators which generates the short
representations are $\O _p \sim [\Tr _G(\phi ^p)]_{(0,p,0)}$, where
$\phi$ is the ${\cal N}=4$ scalar in the adjoint of the gauge group
$G$ and the ${\bf 6}=(0,1,0)$ of the $SU(4)_R$ global symmetry and
$(0,p,0)$ are the Dynkin labels of the $SU(4)_R$ representation.  In
addition, there are multi-trace short representations with primary
operator $\O ^{(p_i)}_p=[\prod _i \Tr _G (\phi ^{p_i})]_{(0,p,0)}$,
with $p=\sum _i p_i$ \WS.  These primary operators all have dimension
$\Delta =p$, independent of $g_{YM}$.  As mentioned in the
introduction, we won't bother distinguishing between single and
multi-trace operators with the same quantum numbers.  The descendent
$Q ^n\overline Q^m \O _p$ has dimension $p+\half (n+m)$.

The short multiplet associated with the primary operator $\O _2\sim
[\Tr _G (\phi ^2)]_{(0,2,0)}$ is very special.  It contains the
conserved currents: the stress-tensor, $T_{\mu \nu}\sim Q ^2\overline
Q ^2 \O _2$, $SU(4)_R$ currents, $J^a_\mu \sim Q \overline Q \O _2$,
and supercurrents $j_{\alpha \mu}^I\sim Q ^2\overline Q \O _2$ and
$\overline j _{\dot \alpha \mu I}\sim \overline Q^2Q\O _2$.

In addition, $\O _2$ contains the exactly marginal scalar operators
$\Ot
\sim Q ^4 \O _2$ and $\Otbar \sim \overline Q ^4 \O _2$.  These
operators can be added to the Lagrangian density without breaking any
supersymmetries, as $Q$ and $\overline Q $ annihilate them up
to total derivatives.  Indeed, adding $\delta \tau \Ot$ to the
Lagrangian changes the gauge coupling $\tau \rightarrow \tau + \delta
\tau$, which preserves all the supersymmetries.  Adding $\Ot$ also
exactly preserves conformal invariance, and thus the full
superconformal group, as $\Ot$ has dimension exactly 4 and is thus
exactly marginal.  This corresponds to the fact that the ${\cal N}=4$
super-Yang-Mills theory has exactly vanishing beta function for all
$\tau$, $\beta _\tau (\tau )\equiv 0$.

Now consider a non-conformal theory, with 16 conserved
supercharges.  Such a theory can be obtained by starting, for example,
with the conformal Lagrangian density ${\cal L}_0$ and adding terms 
which are annihilated, up to total derivatives, by the supercharges
$Q$ and $\overline Q$:
\eqn\Ldef{{\cal L}={\cal L}_0+\sum _p h_pQ ^4\overline Q^4
\O_p +\sum _i h_i^{(L)}Q ^8\overline Q ^8 \O _i^{(L)},}
where $\O _p$ are the short primary operators, including multi-trace
operators, and $\O _i^{(L)}$ are generic long primary operators.  The
$h_p$ and $h_i^{(L)}$ are some real coupling constant parameters\foot{
The notation in \Ldef\ is slightly deceptive: the supercharges
$Q$ and $\overline Q$ will themselves depend on the
parameters $h$, so the Lagrangian is not necessarily linear in the
parameters $h$.  One can find the form of the Lagrangian via an
iterative procedure, first taking the $Q$ in \Ldef\ to be
independent of the $h$, then find the first order in $h$ correction to
$Q$ and plug back in
\Ldef\ to find the next order correction to $Q$, and so on, until
one obtains a Lagrangian which is invariant under its conserved
supercharges.  This is similar to the situation in ${\cal N}=1$
supersymmetry, where one adds a superpotential $\delta {\cal L}=Q ^2
W+c.c.$.  The above iterative procedure properly leads to a Lagrangian
containing both the linear term $W''\psi \psi$ and the quadratic term
$\sim |W'|^2$.}, which we will often refer to collectively as $h$, as
in figure 1.  The short multiplet operator $Q ^4 \overline Q ^4 \O_p$
is in the $(0,p-4,0)$ representation of $SU(4)_R$ (it vanishes, up to
total derivatives, for $p<4$).  The operator $\O _H\equiv Q^4\overline
Q ^4 \O _{p=4}$ is the least irrelevant such perturbation and is an
$SU(4)_R$ singlet.  Again, the short operator $\O _{p=4}$ here can
actually be any combination of the single trace operator $[\Tr _G(\phi
^4)]_{(0,4,0)}$ and the double trace operator $[\Tr _G(\phi ^2)\Tr
_G(\phi ^2)]_{(0,4,0)}$.  The deformation \Ldef\ by the short
operators $Q ^4 \overline Q ^4 \O_{p>4}$ break $SU(4)_R$ for $p>4$.
There are many long operators $Q ^8\overline Q ^8\O _i^{(L)}$ which
preserve the $SU(4)_R$ symmetry, for example there are $SU(4)_R$
singlets in $Q^8\overline Q^8[\Tr _G(\phi ^p)]_{(0,0,0)}$, as well as
many which break $SU(4)_R$.

The deformed theory \Ldef\ again has short multiplets, with primary
operator $\O _p$ in the $(0,p,0)$ representation\foot{We can discuss
the $SU(4)_R$ transformation properties of operators even if \Ldef\
violates $SU(4)_R$.  We can think of the parameters $h$ in \Ldef\ as
expectation values of background fields which carry charge under
$SU(4)_R$ so that $\Ldef$ is invariant, replacing explicit $SU(4)_R$
breaking with spontaneous breaking, much as in \NSback.  $SU(4)_R$
thus still constrains the theory and can lead to selection rules for
the $h$.}  of $SU(4)_R$ given by the same single or multi-trace
operators involving $\phi ^p$.  These short multiplets again consist
of operators $Q ^n \overline Q ^m \O _p$, with $n\leq 4$ and $m\leq
4$, and with $Q$ or $\overline Q$ acting on $Q ^4\overline Q^4 \O_p$
vanishing up to total derivatives.  Indeed, this is necessary for
\Ldef\ to preserve the 16 supersymmetries (see footnote 1).  This is
the ${\cal N}=4$ extension of the statement in ${\cal N}=1$
supersymmetry that chiral operators form short supersymmetry
multiplets, whether or not the theory is conformally invariant.

Let's initially restrict our attention to theories \Ldef\ which
preserve the $SU(4)_R$ symmetry.  It is then clear that all of the
$(p+2)^2((p+2)^2-1)/12$ operators in the primary operator $\O _p$
(this is the dimension of the $(0,p,0)$ representation of $SU(4)_R$)
must have the same dimension $\Delta _p$.  A' priori, as the theory is
not conformally invariant, $\Delta _p$ can depend on the RG scale
$\mu$, as well as the gauge group, $g_{YM}$, and the parameters $h$ in
$\Ldef$.  However, we will now argue that this is not the case.

In terms of ${\cal N}=1$ supersymmetry, the real scalars $\phi$ in the
$(0,1,0)$ representation of $SU(4)_R$ are the bosonic components of
chiral superfields $\Phi$ in the $3_{2/3}$ of $SU(3)_F\times
U(1)_R\subset SU(4)_R$ and anti-chiral superfields $\overline \Phi$ in
the $\overline 3_{-2/3}$.  The operators $\O_p$ then contains the
bosonic component of purely chiral superfields $\Tr \Phi ^p$ in the
$[(p+2)(p+1)/2]_{2p/3}$ representation of $SU(3)_F\times U(1)_R$,
along with the conjugate, purely anti-chiral superfields, and mixed
operators, which contain both chiral and anti-chiral superfields.  We
now use the fact that ${\cal N}=1$ supersymmetry implies that the
purely chiral superfields form a ``chiral ring,'' with purely additive
anomalous dimensions at all scales.  Thus the purely chiral
superfields $\Tr \Phi ^p$ must have dimensions $\Delta _p=p\Delta _1$
for some $\Delta _1$ which, a' priori, might still depend on the RG
scale.

By the $SU(4)_R$ symmetry, all operators in $\O _p$, in
the $(0,p,0)$ $SU(4)_R$ representation, must then have dimension
$\Delta _p=p\Delta _1$ for all RG scales and coupling constant
parameters.   Similarly, using the supersymmetry algebra, the
supersymmetry descendents $Q ^n \overline Q^m \O _p$
must have dimension $\half (n+m)+p\Delta _1$ for all RG scales and
coupling constant parameters.  

Even though the theory is not conformally invariant, the conserved
currents are given by the supersymmetry descendents of the $\O_2$
operators.  As in the conformal theory, the conserved stress tensor is
$Q ^2 \overline Q^2 \O_2$ and the conserved supercurrents
are $Q^2 \overline Q \O_2$ and $Q\overline Q ^2
\O_2$.  By the discussion above, the stress tensor $T_{\mu \nu}$ 
will thus have dimension $2+2\Delta _1$.  But, because $T_{\mu \nu}$
is a conserved current, its scaling dimension can not be renormalized:
it must be exactly 4.  Thus $\Delta _1 \equiv 1$ and the dimensions of
all short operators are not renormalized: $Q ^n \overline Q
^m \O_p$ has dimension exactly given by $p+\half (n+m)$ for all RG
scales and coupling constant parameters.  

In particular, the operator $\Ot \sim Q ^4 \O_2$, which can be
added to the Lagrangian to change the gauge coupling $\tau$, has
dimension exactly 4 for all RG scales and coupling constant
parameters.  Thus $\delta {\cal L}\sim \delta \tau \int d^4 x \Ot$ is
an exactly marginal deformation for all scales and coupling constants,
which implies that the gauge coupling beta function continues to
vanish identically, $\beta _\tau
(\tau, h)\equiv 0$, even though the theory is not conformally
invariant.  The gauge coupling $\tau$ does not vary with the RG scale,
as depicted in figure 1. 

We now relax our restriction that the deformation \Ldef\ preserves the
$SU(4)_R$ symmetry.  One might then worry that the dimensions of the
different operators in $\O _p$, which were previously equal as they
were in the irreducible $(0,p,0)$ representation of $SU(4)$, could now
be split.  But any possible splitting must be compatible with imposing
the $SU(4)_R$ symmetry which can be regarded as being unbroken
provided the parameters $h$ in \Ldef\ are assigned appropriate
$SU(4)_R$ transformation properties.  The only natural choice is to
have the possible splitting of the operator dimensions be a
combination of the parameters $h$ which is in the $(0,p,0)$
representation of $SU(r)_R$.  Similarly, the possible splittings of
the operators in $Q\O _p$ should be in the same representation as this
collection of operators, the $(0,p-1,1)$ representation of $SU(4)_R$.
But, for any operator in $\O _p$ with dimension $\Delta$, the
corresponding operator in $Q\O _p$ must have dimension $\Delta +
\half$; this would not be compatible with having non-trivial
splittings of the $\O _p$ operators dimensions in the $(0,p,0)$ of
$SU(4)_R$ and $Q\O _p$ in the $(0,p-1,1)$.   We thus conclude
that all operators in $Q^n \overline Q ^m \O _p$ 
which are related by $SU(4)_R$ rotations must have the same operator
dimensions, even if \Ldef\ breaks $SU(4)_R$.   The above argument that
these operators all must then have unrenormalized dimensions, for all
RG scales and coupling constant parameters, then goes through exactly
as in the case where \Ldef\ respects $SU(4)_R$.  

Note that it is necessary to have 16 preserved supersymmetries for the
above arguments to go through.  For example, if ${\cal N}=4$ is broken
to ${\cal N}=2$ (via adjoint masses) the gauge beta function no longer
vanishes.  This can happen because the operator $\Ot$ associated with
changing the gauge coupling is no longer in the same supermultiplet as
the conserved stress tensor $T_{\mu \nu}$, and can thus obtain an
anomalous dimension.  

\nref\MDNS{M. Dine and N. Seiberg, hep-th/9705057, 
\pl{409}{1997}{239}.}
\nref\PSS{S. Paban, S. Sethi, M. Stern, hep-th/9805018,
\np{534}{1998}{137}; S. Sethi and M. Stern, hep-th/9903049, JHEP
9906:004 (1999).}
\nref\DLRU{F. Gonzalez-Rey, B. Kulik, I.Y. Park, M. Rocek,
hep-th/9810152, \np{544}{1999}{218}; E.I. Buchbinder, I.L. Buchbinder,
S.M. Kuzenko, hep-th/9810239, \pl{446}{1999}{216}; D.A. Lowe, R. von
Unge, hep-th/9811017, JHEP 9811 (1998) 014.}
\nref\MDJG{M. Dine and J. Gray, hep-th/9909020.}
\def\nonrefs{\refs{\MDNS - \MDJG}}
Finally, the reader might wonder about any possible connections
between the non-renormalization of operator dimensions discussed here
and other non-renormalization theorems concerning ${\cal N}=4$
theories, which have been of recent interest \nonrefs\ in the context
of the matrix description of $M$ theory.  The non-renormalization
theorems of \nonrefs\ concern the effective action for the pure ${\cal
N}=4$ super-Yang-Mills theory along the Coulomb branch, where the
unbroken gauge group is generically the Abelian Cartan torus $U(1)^r$
of the gauge group.  The low-energy theory is then ${\cal N}=4$, with
$U(1)^r$ gauge group and 16 supersymmetry preserving deformations, of
the form
\Ldef, which arise via integrating out the massive gauge bosons of
the original gauge group.  In this situation, where the scalar
expectation values are generic and the low-energy gauge group is
purely abelian, the IR fixed point is a free-field theory, which is
perturbed by the irrelevant operators in \Ldef.  

Because the IR theory is free in this case, the statement of the
present paper, that certain operator dimensions are not renormalized,
is essentially trivial in this context.  The non-trivial
non-renormalization theorems of \nonrefs, in the context of the IR
free theories, concerns the exact expressions for the {\it
coefficients} $h$ appearing in \Ldef.  Perhaps it is also possible to
obtain exact expressions for some of the coefficients $h$, appearing
in a more general effective action of the form \Ldef, even in our
present context, where the IR theory is non-Abelian, and thus not
free.

\newsec{Holography and RG flows}

A motivation for considering 4d RG flows with 16 preserved
supersymmetries comes {}from the AdS/CFT correspondence \review\ and
some conjectured extensions.  Recall that type IIB string theory has
vacua with metric
\eqn\metric{ds^2=H^{-1/2}\eta _{\mu \nu}dx^\mu dx ^\nu +H^{1/2}
\delta _{ab}dy^a dy^b,}
with coordinates $x^\mu$ spanning $\IR ^{1,3}$ and $y^a$, which we
often write as $\vec y$, spanning $\IR ^6$.  $H(y)$ is an arbitrary
function which satisfies the $y^a$ Laplace equation, allowing for
possible delta function source terms: $\Delta _y H=-2\pi ^{1/2}\kappa
\sum _{i=1}^N \delta ^{6}(\vec y-\vec a_i)$, where the $\vec a_i$ are
arbitrary vectors in $\IR ^6$.  These vacua have $F_5\sim \kappa
^{-1}(1+*)(dx)^{\wedge 4}\wedge dH^{-1}$, with $N$ units of integrated
$F_5$ flux.  They generally preserve 16 supersymmetries.  A point
which we would especially like to emphasize is that these vacua all
have {\it constant} axio-dilaton $\tau$, for arbitrary $\tau$ in the
upper half plane.  We will connect this with the statement in the
previous section that $\beta _{\tau}\equiv 0$ with 16 supersymmetries.

A special case, which preserves the $SO(6)\cong SU(4)$ symmetry of
rotating $\vec y$, is
\eqn\Hspec{H=h+{R^4\over r^4},}
where $h$ is an arbitrary real constant, $R^4\equiv 4\pi g_s\alpha
^{\prime 2}N\equiv N\kappa /2\pi ^{5/2}$, and $r^2=\delta
_{ab}y^ay^b\equiv \vec y\cdot
\vec y$.  Choosing the constant $h=1$, the metric is the $D3$ brane
metric \HorStrom, which asymptotes for $r\rightarrow
\infty$ to flat, 10 dimensional space-time.  For $r\rightarrow 0$,
there is a tube in $r$ with an $S^5$ section of radius $R$ and
non-singular horizon at $r=0$.  Rather than choosing $h=1$, one could
take $h=0$, which is the $AdS_5\times S^5$ vacuum of IIB string
theory.  The two cases, $h=1$ and $h=0$, are clearly asymptotically
the same in the near horizon limit, $r\rightarrow 0$, but differ
drastically for $r\rightarrow \infty$.  While the $h=1$ case
approaches flat 10d space-time in this limit, the $h=0$ case always
remains $AdS_5\times S_5$, with the $AdS_5$ coordinate $r$ approaching
a boundary which $AdS_5$ has at $r\rightarrow \infty$.  It is this
boundary (rather than the horizon) where operators are inserted in the
prescription of \refs{\GKP,\EW}.

The $h=1$ case preserves 16 supersymmetries for generic $r$ and
asymptotically preserves an additional 16 supersymmetries in the
limits $r\rightarrow 0$ and $r\rightarrow \infty$. In the
$r\rightarrow 0$ limit, the symmetries combine into the supergroup
$PSU(2,2|4)$, which is the 4d ${\cal N}=4$ superconformal group.  The
$h=0$ case, $AdS_5\times S^5$, identically has the $PSU(2,2|4)$
superconformal symmetry group, with the 32 supersymmetries, for all
$r$.

The works \refs{\kleba, \GKT, \SGIK} considered the $h=1$ case and
compared the asymptotic scattering of bulk waves from $r\rightarrow
\infty$ to the dynamics of the world-volume gauge theory of the D3
brane.  Similarly, \malda\ started from the $h=1$ case and then
conjectured duality with the world-volume gauge theory in the
$r\rightarrow 0$ near-horizon limit.

By starting instead with the $h=0$ case, it is not necessary to take
the near horizon limit: the $h=0$ case is conjectured to be
holographically dual, for all energies and for all $r$, to ${\cal
N}=4$ super-Yang-Mills, with gauge group $SU(N)$.  The fact that the
dilaton is an arbitrary constant $\tau$ corresponds to the fact that
the ${\cal N}=4$ Super-Yang-Mills theory is exactly conformally
invariant for arbitrary $\tau$.  The fact that the $h=0$ case has
symmetry group $PSU(2,2|4)$ for all $r$ shows that the holographic
dual is {\it precisely} the ${\cal N}=4$ super-Yang-Mills theory and
not, for example, some deformation of this theory, such as
Born-Infeld, which would not be exactly conformally invariant. 

On the other hand, following \refs{\GHKK, \SGAH}\ we would like to
interpret the case with $h\neq 0$ as being holographically dual to a
deformation of ${\cal N}=4$ super-Yang-Mills.  Interpreting $r$ as the
renormalization group energy scale parameter, this deformation flows
to the ${\cal N}=4$ superconformal fixed point in the infrared,
$r\rightarrow 0$, as $h$ becomes irrelevant in this limit.  The
deformation of ${\cal N}=4$ is thus by operators which become
irrelevant in the IR.  In \GHKK\ the deformation was regarded as the
replacement of the ${\cal N}=4$ Yang-Mills Lagrangian with a
Born-Infeld generalization (though no such non-Abelian ${\cal N}=4$
Born-Infeld action is known as of yet).  A simpler possibility for the
IR irrelevant deformation was conjectured in \SGAH\ on the basis of
$PSU(2,2|4)$ representation theory.  This will be discussed further in
what follows. (We also mention the possibility that the seemingly
simpler deformation actually is ${\cal N}=4$ Born Infeld.) 

The more general vacuum solution is given by \metric\ with 
\eqn\Hgen{H(y)=h+4\pi g_s\alpha ^{\prime 2}\sum _{i=1}^k{N_i\over
|\vec y-\vec a_i|^4},} for some integers $N_i$ and vectors $\vec a_i$
in $\IR ^6$.  For $h=1$ this corresponds to separating $N=\sum _i N_i$
$D3$ branes, placing $N_i$ at $\vec y = \vec a_i$.  We can choose the
origin of $\vec y$ so that $\sum _i N_i \vec a_i =0$.  The general
solution \Hgen\ does not preserve $SU(4)_R$ but, again, does preserve
16 supersymmetries.  Also, we again emphasize that the dilaton is
constant with the general solution \Hgen.  

We note that the metric \metric, with \Hspec, is invariant under the
transformation:
\eqn\RGflow{x_{\mu }\rightarrow \lambda
x_{\mu }, \quad r\rightarrow \lambda ^{-1} r,  
\quad h\rightarrow \lambda ^4h, \quad d\Omega
_5\rightarrow d\Omega _5,} with $\lambda$ an arbitrary real parameter
and $g_s$ and $\alpha '$ held fixed.  Similarly, the more general
solution \Hgen\ is invariant under the transformation
\eqn\RGfloww{x_{\mu }\rightarrow \lambda
x_{\mu }, \quad \vec y \rightarrow \lambda ^{-1}\vec y, 
\quad  \quad h\rightarrow \lambda ^4h, \quad \vec a_i 
\rightarrow \lambda ^{-1}\vec a_i.}  
This transformation leaves $F_5\sim \kappa ^{-1}(1+*)(dx)^{\wedge
4}\wedge dH^{-1}$ invariant, so it preserves the $F_5$ flux
quantization condition.  This transformation is a symmetry of the
full, interacting, IIB supergravity, with the various fields,
e.g. $F_5$, the anti-symmetric $B_{\mu \nu}$ fields and, in
particular, the dilaton, all invariant.

We expect that the above transformation is an exact symmetry of the
full IIB string theory in these vacua with $F_5$ flux.  For $h=0$,
this symmetry is the exact gauge symmetry of the string theory vacuum
under the dilatation element of the symmetry group $PSU(2,2|4)$.  It
is on a similar footing as the full translational invariance of the
flat $\IR ^{1,9}$ vacuum of IIB string theory.  In the field theory
dual, the transformation
\RGflow\ or \RGfloww\ are interpreted as the renormalization group
flow, with $\lambda \rightarrow 0$ in the UV and $\lambda \rightarrow
\infty$ in the IR. 
The fact that $\vec y$ or $r$ scale oppositely from $x_\mu$ is the
well-known UV-IR correspondence \SW.  The fact that $g_s=g_{YM}^2$ is 
invariant under this transformation corresponds to the fact that the
Yang-Mills gauge coupling beta function vanishes identically along RG
flows. 

For $h=0$ and all $\vec a_i =0$, the invariance under \RGflow\
corresponds to the fact that the dual theory is conformally invariant
and thus unchanged by RG flow.  For $h\neq 0$, this symmetry is
broken, but can be regarded as being restored provided $h$ transforms
as in \RGflow.  In other words, the transformation of $h$ as in
\RGflow\ exactly compensates for the fact that the theory is not
invariant under scale transformations, with the theory unchanged under
the combined transformation \RGflow.  This means that, at length
scales $x$, the theory only depends on $h$ via the effective coupling
$h_{eff}\sim h/x^4$, which is invariant under \RGflow.   
Similarly, for $\vec a_i\neq 0$, the transformation in
\RGfloww\ indicates that the theory only depends on this parameter via
the invariant combination $\vec a_{i,eff}\sim \vec a_ix$.

We interpret the power of $x$ in $h_{eff}\sim h/x^4$ as showing that,
in the holographic dual 4d field theory, $h$ is a coupling constant
which multiplies a term of energy dimension {\it exactly} 4 in the
action $S$, i.e. a term of energy dimension {\it exactly} 8 in the
Lagrangian density ${\cal L}$. For any fixed $h$, $h_{eff}$ indeed
vanishes at long distances, $x\rightarrow
\infty$, which is the statement that the theory flows to the conformal
${\cal N}=4$ RG fixed point in the far IR.  Similarly, the parameters
$\vec a_i$ correspond to scalar $\vec \phi$ expectation, which Higgs
$SU(N)\rightarrow S(\prod _i U(N_i))$. The fact that the theory only
depends on $\vec a_{i,eff}\sim \vec a_ix$ is simply the statement that
$\vec \phi$ has exact energy dimension 1.

We now argue that the theory actually only depends on $h$ via the
dimensionless quantity $h_{eff}=hR^4/x^4$.  Consider the scale
transformation
\eqn\hRscale{x^\mu \rightarrow x^\mu, \quad g_s\rightarrow g_s, \quad
h\rightarrow \lambda ^{-4}
h, \quad \alpha '\rightarrow \lambda ^2 \alpha ', 
\quad \vec y\rightarrow
\lambda ^{2}\vec y, \quad \vec a _i \rightarrow \lambda ^2\vec a_i.}
Unlike \RGflow\ and \RGfloww, the metric \metric\ is not invariant
under \hRscale, but rather scales as $ds^2 \rightarrow \lambda ^{2}
ds^2$; also unlike \RGflow\ and \RGfloww, the string tension 
(and thus the gravitational coupling $\kappa$) scales under
\hRscale.  Since $\alpha ' \sim (length)^2$ and $ds^2 \sim (length)^2$,
the transformation in \hRscale\ amounts to a rescaling of all lengths
by a factor of $\lambda$.  Since all lengths in string theory are
measured relative to $\alpha '$, IIB string theory and all of its
scattering amplitudes must be invariant under
\hRscale.  Because the absorption probability for waves scattered from
$r=\infty$ to $r=0$ do not depend on $r$, the scaling of $r$ in
\hRscale\ is immaterial.  The essential point is that the absorption
probability only depends on $h$ via the dimensionless quantity
$h_{eff}=hR^4/x^4$, which is invariant under \RGflow, \RGfloww, and
\hRscale.  We have chosen to write $h_{eff}$ in terms of $R$ rather
than $\alpha '$ because the leading order supergravity results depend
on $\alpha '$, $g_s$, and $N$ via the combination $R^4\equiv 4\pi
g_s\alpha ^{\prime 2}N$.  Away from the leading order supergravity
limit, there can be additional explicit dependence on $g_s$ and $N$.

Similarly, $\vec a_{i, eff}=\vec a_i x/R^2$ is the quantity
which is invariant under \RGfloww\ and \hRscale.  

So the conjecture is that IIB string theory in the vacuum \metric\
with \Hspec, for general $h$, is holographically dual to the 4d field
theory with Lagrangian
\eqn\Ldefspec{{\cal L}={\cal L}_0+ (const.)\ h R^4\O _H,}
where ${\cal L}_0$ is the Lagrangian of the ${\cal N}=4$
superconformal theory, with gauge group $G=SU(N)$, and $\O_H$ is an
operator, of dimension exactly 8, which preserves 16 supersymmetries,
and which preserves the $SU(4)_R$ symmetry.  $\O_H$ is thus $Q ^4
\overline Q^4 \O _{p=4}$, where $\O _{p=4}$ can, again, be a
combination of $[\Tr _G(\phi ^4)]_{(0,4,0)}$ and $[\Tr _G(\phi ^2)\Tr
_G(\phi ^2)]_{(0,4,0)}$.  For $h=1$, this was originally conjectured
in \SGAH\ based on the fact that $\O _H$ is the unique scalar
$SU(4)_R$ singlet (besides $\Ot$ and $\Otbar$) in the short
representations of $PSU(2,2|4)$.

As mentioned in footnote 1, the deformation \Ldefspec\ is actually
not linear in $h$, as the supercharges in $O_H=Q^4\overline Q^4 \O _4$
get $h$ dependent corrections from \Ldefspec.  It is interesting to
speculate that perhaps \Ldefspec\ actually generates the full ${\cal
N}=4$ Born Infeld Lagrangian.

In line with our discussion in the previous section, the gauge
coupling should be an arbitrary constant, for all RG scales, with the
Lagrangian \Ldefspec.  This is clearly the case in the holographic
gravity dual, as the axio-dilaton is an arbitrary constant with the
vacuum \metric\ and \Hspec\ or \Hgen.  In particular, it does not vary
with the coordinate $r$, which corresponds to the RG scale.  Also,
based on \RGflow, we argued that $h$ should multiply an operator of
dimension exactly 8 in the Lagrangian, and this should be the case for
arbitrary $h$.  This is in line with our argument in the previous
section that the dimensions of operators in short representations are
not modified by deformations \Ldef, of which \Ldefspec\ is a special
case.

While the dimension 8 operator $\O_H$ is the unique short
representation operator which could enter in the conjectured field
theory dual
\Ldefspec\ of IIB string theory with \metric\ and \Hspec, there could
also be long operators, as in \Ldef, which preserve the 16
supersymmetries and $SU(4)_R$.  Consider such an operator $Q^8
\overline Q^8 \O_{long}$, of operator dimension $\Delta
(g_{YM})$, which generally depends on $g_{YM}$.  To preserve the
invariance under \RGflow, the parameter $h$ would have to multiply
such a term in the Lagrangian with a power $h^{(\Delta (g) -4)/4}$.
This $g_{YM}$ dependent power of $h$ would lead to a non-trivial
$g_{YM}$ RG running, in contrast to the fact that the axio-dilaton
remains constant.  It thus seems likely that \Ldefspec\ is actually
the {\it exact}\foot{However, \Ldefspec\ leads to contact terms whose
cancellation should require additional counter-terms (e.g. $\sim (h
\O _H)^n$); I thank G. Moore and S. Shatashvili for stressing this
point to me.}  holographic dual of IIB string theory with \Hspec,
without any additional short or long operators, for any $g_{YM}$ and
$N$.

The above, conjectured, duality between the IIB string theory with
\Hspec\ and the field theory with \Ldefspec\ is conjectured to hold
for all RG scales.  In particular, the UV limit of this flow
corresponds to $r\rightarrow \infty$ and thus gives a field theory
which is holographically dual to the asymptotically flat
10-dimensional, Minkowski space-time!  Note that we are defining the
UV limit of a RG flow in terms of an irrelevant perturbation of the IR
fixed point, i.e. defining the UV theory via reversing the RG flow.
This requires fine-tuning the coefficients of every irrelevant
operator so that the Lagrangian is exactly \Ldefspec\ at all scales.

As already mentioned, the invariance \RGflow\ is to be interpreted as
RG flow in the dual field theory.  The correlation functions of
operators in the dual field theory thus satisfy a corresponding Ward
identity
\eqn\RGward{\left(4h{\partial\over \partial h}+\sum _{i=1}^n 
(x_i{\partial \over \partial x_i}+\Delta _i)\right)\ev{\prod
_{i=1}^n\O _i(x_i)}_h=0,} where $\Delta _i$ are the operator
dimensions and the $h$ subscript is a reminder that the expectation
value is in the theory deformed by the parameter $h$.  Using the fact
that the Lagrangian is \Ldefspec, with the $h$ term the only part
which violates scale invariance, we have, $T_\mu ^\mu \sim h R^4 \O
_H$, Thus, for arbitrary correlation functions,
\eqn\hderiv{4h{\partial \over \partial h}\ev{\prod _{i=1}^n \O _i
(x_i)}_h =-\int d^4 y\ev{T_\mu ^\mu (y)\prod _{i=1}^n \O _i
(x_i)}_h.}
It then follows that \RGward\ is the Callan-Symanzik Ward
identity for correlation functions in a theory which is not
conformally invariant,
\eqn\callanw{\sum _{i=1}^n (x_i
{\partial \over \partial x_i}+\Delta
_i)\ev{\prod _{i=1}^n\O _i (x_i)}_h=
\int d^4 y\ev{T_\mu ^\mu (y)\prod _{i=1}^n \O _i
(x_i)}_h.}
By the argument of the previous section, the dimensions $\Delta _i$
of operators in short representations are not renormalized and are
independent of $h$.  

In particular, \RGward\ implies that 2-point functions are of the form
\eqn\tpfgen{\ev{\O _{\Delta}(x)\O _{\Delta}(0)}_h={c_{\Delta}
\over x^{2\Delta}}f_{\Delta}(h_{eff}, g_{YM}, N),}
where $c_{\Delta}$ is a h-independent constant, $h_{eff}=hR^4/x^4$,
and $f_{\Delta}(h_{eff})$ is a function which can be normalized to
equal 1 for $h_{eff}=0$, which is the far IR limit.  For long
operators, the dimension $\Delta $ appearing in the exponent in
\tpfgen\ could also be a function $\Delta (h_{eff}, g_{YM}, N)$.
However, for short operators, as we have argued in the previous
section, the dimension $\Delta$ is an unrenormalized constant,
independent of $g_{YM}$, $h$, and the RG scale.  For example, the
2-point function of $\Ot$ with its conjugate operator $\Otbar$ should
be given by \tpfgen\ with $\Delta
\equiv 4$.

This can be compared with the calculation of \SGAH, where the dilaton
2-point function was computed in the theory with $h=1$ via a
supergravity computation of the corresponding partial-wave absorption
cross section.  Restoring the $h$ dependence via the argument which
follows \hRscale, the result of \SGAH\ is\foot{ As discussed in \KI\
all leading supergravity results for correlation functions are
proportional to $\hbar ^{-1}\sim R^3\kappa _5^{-2}\sim N^2-1$, with
the replacement of $N^2$ by $N^2-1$ presumably coming from a one-loop
string correction to the relation between $\kappa _{10}$ and $\kappa
_5$ in $S^5$ dimensional reduction.}
\eqn\GHres{\ev{\Ot (x)\Otbar (0)}= {3(N^2-1)\over \pi ^4
x^8}f({hR^4\over x^4}),} for a function $f(t)$ which was determined in
terms of solutions of Mathieu's equation.  Note, in particular, that
this is indeed of the form \tpfgen, with the $x$ exponent, $2\Delta =
8$, an unrenormalized constant and not a non-trivial function $\Delta
(h_{eff})$ of $h_{eff}=hR^4/x^4$.

So the result of \SGAH\ is compatible with our statement that the
dimension of $\Ot$ is not renormalized.  A scale-dependent
renormalization of the dimension of $\Ot$ was obtained in \SGAH\
because the right hand side of \callanw, involving $T_\mu ^\mu$, was
omitted.  Again, the fact that the dimension of $\Ot$ is precisely 4,
for all RG scales and $h$, agrees with the fact that the gauge
coupling is an arbitrary constant, which does not change with the RG
scale.  This agrees with the fact that the dilaton is constant,
independent of $r$ and $x_\mu$, for the vacuum \Hspec.

The function $f(t)$ of \GHres\ is given by \SGAH\ as
\eqn\ffunctis{f(t)= \sum _{n=0}^\infty 
\sum _{k=0}^n c_{n,k}t^n(\half \log
t)^k,} with coefficients $c_{n,k}$ which are independent of $g_{YM}$
and $N$, at least in the large $N$ limit, and given in \SGAH\ via a
complicated expression.  The first few terms quoted in \SGAH\ are
\eqn\cquote{c_{0,0}=1\qquad \eqalign{c_{1,1}&=-320\cr
c_{1,0}&=-1024}\qquad \eqalign{c_{2,2}&=571200\cr c_{2,1}&=4408560\cr 
c_{2,0}&=\f{14}{3}(1422697-12000\pi ^2).}}
This can be compared with \Ldefspec, which gives
\eqn\cdefi{\ev{\prod _i \O _i (x _i)}_h 
=\ev{e^{-(const.)\ hR^4 \int d^4 y
\O _H(y)}\prod _i \O _i (x _i)} _{h=0}.}  The constant appearing in 
\Ldefspec\ and \cdefi\ can be fit so that \cdefi\ reproduces, say
$c_{1,1}$.  The rest of the $c_{n,k}$ appearing in the function $f(t)$
of \GHres\ should then be completely determined by
\cdefi; it would be interesting to complete this check.  Perhaps the 
equations of \SGAH\ for the function $f(t)$ can be obtained directly
using \hderiv\ with $T_\mu ^\mu \sim hR^4\O _H$.

The above considerations can be similarly applied for the more general
vacuum \Hgen, which corresponds to the theory with Lagrangian
\Ldefspec, deformed away from the origin of the moduli space of vacua,
where $SU(N)\rightarrow S(\prod _i U(N_i))$.  

As a particular example, consider the theory with $H=4\pi g_s\alpha
^{\prime 2}((N-1)/|\vec y|^4+1/|\vec y - \vec a|^4)$, where we shifted
the $\vec y$ origin for convenience.  This is expected to be dual to
the RG flow from the ${\cal N}=4$ superconformal theory with gauge
group $SU(N)$ in the UV and $SU(N-1)\times U(1)$ in the IR (large
$\vec y$ is the UV and small $\vec y$ is the IR).  The least
irrelevant operator IR, along which the theory flows to the IR fixed
point, is $\O _H\sim Q^4\overline Q^4 \O _4$.  The coefficient of this
operator is $\sim 1/v^4$, where $v$ is the Higgs expectation value
which breaks $SU(N)$ to $SU(N-1)\times U(1)$.  We can also see this
{}from the above $H(\vec y)$, which is given by $H(y)\approx H=4\pi
g_s\alpha ^{\prime 2}((N-1)/|\vec y|^4+1/|\vec a|^4)+\dots $ for small
$\vec y$.  The IR theory thus effectively has $H(y)$ given by 
\Hspec\ with $h=R^4/N|\vec a|^4$.  If
we identify $v\sim a/R^2$, the coefficient $hR^4$ of $\O _H$ is indeed
$1/v^4$.  Using the analysis of \IKEW\ it should be possible to find
the precise relation between $v$ and $a/R^2$ and, by comparing with
the precise coefficient of $\O _H$ induced by the above Higgsing, 
obtain an independent derivation of the constant appearing in
\Ldefspec.  It would be interesting to complete this exercise and to 
compare the value of the constant thus obtained with that required to
reproduce \ffunctis\ and \cquote\ via \cdefi.

\newsec{Three dimensional theories with 16 or 32 supercharges}

We now consider three dimensional gauge theories with 16 supercharges,
which is sometimes referred to as ${\cal N}=8$ supersymmetry in 3d.
Useful aspects of these theories can be found in \sixteen.  The
supercharges are 8 $SO(2,1)$ spinors $Q^I_\alpha$, where $I=1\dots 8$
and $\alpha =1,2$ is the Lorentz spinor index.  The supersymmetry
algebra admits an $SO(8)_R$ automorphism, with $I$ taken to reside in
the ${\bf 8}_s$; the Yang-Mills Lagrangian is invariant only under an
$SO(7)_R$ subgroup for general gauge coupling $g_{YM}$.  The three
dimensional theory can be regarded as the dimensional reduction of the
4d ${\cal N}=4$ theory on a circle of radius $R\rightarrow 0$.  The
$SU(4)_R\cong SO(6)_R$ of the 4d theory then extends to the $SO(7)_R$
symmetry of the Lagrangian and the $SO(8)_R$ symmetry of the
supersymmetry algebra; under $SO(6)_R\subset SO(7)_R\subset SO(8)_R$,
the supercharges combine as ${\bf 4}+\overline{\bf 4}\rightarrow {\bf
8}_s\rightarrow {\bf 8}_s$.

As opposed to the situation in 4d, in 3d the gauge coupling $g_{YM}$
is classically dimensionful, flowing to strong coupling in the IR.  It
is believed that the coupling flows until it reaches some fixed point
value $g_{YM}^*$ where the beta function vanishes and the theory is
conformally invariant and interacting.  At this point, the theory has
a total of 32 supercharges (the original 16 and 16 additional,
superconformal ones), which combine into the superconformal group
$SO(3,2|8)$.

The superconformal group $SO(3,2|8)$ again has both short and long
representations, with supermultiplets generated by primary operators
$\O _P$ via graded commutators with $Q_{\alpha}^I$, which we again
denote by $Q ^n \O _P$.  Here $n\leq 16$ for the long multiplets and
$n\leq 8$ for the short multiplets.  The extremal cases $Q^{16}
\O _{long}$ or $Q ^8 \O _{short}$ can be added to the Lagrangian
without violating the 16 supersymmetries as, up to total derivatives,
they are annihilated by the supercharges.  These deformations of the
superconformal theory are all irrelevant in the far IR.  

The short multiplets of $O(3,2|8)$ were constructed in \osciiid.  They
are given by scalar primary operators $\O _p\sim \Tr \phi ^p$ of
dimension $\Delta _p=\half p$ and in the $(p,0,0,0)$ representation of
$SO(8)_R$, along with descendents $Q ^n\O _p$, for $n\leq 8$, with
$\Delta =\half p + \half n$ and other quantum numbers as reviewed in
appendix A.  The representations with $p<4$ are shorter than the
generic short representation, as the operators in the table in the
appendix which would otherwise have negative $SO(8)_R$ Dynkin weights
actually vanish.  For example, the case $p=1$ is the singleton
representation, given by scalars $\O _1$, in the ${\bf 8}_v$ of
$SO(8)_R$ with $\Delta = \half$, and fermions $Q \O _1$, in the ${\bf
8}_c$ of $SO(8)_R$ with $\Delta =1$; acting with more powers of $Q$ on
$\O _1$ gives zero (up to total derivatives).  This $p=1$ multiplet is
that of the gauge invariant operators in $U(1)$ gauge theory, with one
of the $8$ scalars in $\O _1$ identified as the dualized photon:
$*dA=d\phi _8$.

As in four dimensions, the $p=2$ short multiplet contains the
conserved currents.  The $SO(8)_R$ currents $J_\mu ^a$ (which may or
may not be conserved) are the descendents $Q ^2 \O _2$, the
supercharges are $Q^3 \O _2$, and the conserved stress tensor $T_{\mu
\nu}$ is $Q ^4 \O_2$.  Because $T_{\mu \nu}$ is definitely a conserved
current, its dimension must always be exactly $\Delta =3$.

Now, as in sect. 2, we consider a deformed theory, as in \Ldef, with
16 supersymmetries preserved.  The dimension of $T_{\mu \nu}$ remains 
exactly $\Delta =3$ and thus the dimension of $\O _2$ remains exactly
1.  As in 4d, the chiral ring structure of additive anomalous
dimensions for chiral superfields then ensures that all short
operators $Q^n \O _p$ continue to have their unrenormalized dimension
$\Delta = \half (p+n)$, even in the deformed theory, with arbitrary
deforming parameters $h$ in \Ldef. 

As an aside, we mention some peculiar aspects of the 3d theories:

\lfm{1.} In 4d, the microscopic, non-Abelian, Yang-Mills fields 
are in the $p=1$ short multiplet representation of the supersymmetry
algebra.  The situation in 3d is different because the 
$p=1$ multiplet contains dualized scalars rather than gauge fields.
While 3d Abelian gauge fields can always be dualized to scalars, it is
not known how to do this for non-Abelian gauge fields.  Note also that
the $p=1$ multiplet in 4d contains $\O _1$ descendents up to $Q^2 \O
_1$ and $\overline Q^2 \O _1$ (which is why this multiplet is
sometimes referred to as the ``doubleton'' rather than ``singleton'' 
multiplet), while the $p=1$ multiplet in 3d only contains descendents
up to $Q \O _1$.  

\lfm{2.} Unlike the situation in 4d, 
the $p=2$ short multiplet in 3d does not contain an operator
associated with changing the gauge coupling $g_{YM}$.  The only
candidate for such an operator would be a Lorentz scalar in $Q^4 \O
_2$, which is the 3d analog of the operators $\Ot$ and $\Otbar$ in 4d.
But this operator vanishes in 3d (up to total derivatives); this can
be seen in Appendix A because the scalar in $Q^4 \O _p$ is in the
$(p-4,0,0,2)$ representation of $SO(8)_R$ and must vanish for $p<4$.
For $p\geq 4$ $Q^4 \O _p$ is not annihilated by the supercharges (up
to total derivatives), so there is no value of $p$ for which it can be
added to the Lagrangian while preserving 16 supersymmetries.

The fact that there is no short multiplet operator associated with
changing the gauge coupling $g_{YM}$ actually prevents a
contradiction.  The RG flow associated with $g_{YM}$ changing with RG
scale, until it hits the RG fixed point value $g_{YM}^*$, is one which
preserves 16 supersymmetries.  Thus, by the argument above, the
dimensions of all operators in short representations are not
renormalized along this RG flow.  But the dimension of the operator
responsible for changing $g_{YM}$ must vary with the RG flow, such
that it is relevant for $g_{YM}\ll g_{YM}^*$ and becomes irrelevant in
the IR, for $g_{YM}=g_{YM}^*$ (the intuition is that the fixed point
at $g_{YM}^*$ is attractive in the IR).  If the operator responsible
for changing $g_{YM}$ were $Q^4 \O _2$, it would always be exactly
marginal, which we know to be untrue even for small $g_{YM}$.

We thus expect (though with some confusion) that the operator
associated with changing $g_{YM}$ is actually a long operator which
preserves 16 supersymmetries.

As another peculiar aside, note that the pseudoscalar operator $Q^2 \O
_2\sim \Tr \psi \psi$ can be added to the Lagrangian without breaking
the 16 supersymmetries.  The reason is that $Q$ acts on this operator
to give the spin $\half$ operator in $Q^3\O _2$, which vanishes up to
total derivatives.  As seen in the table in appendix A, the spin
$\half$ operator in $Q^3 \O _p$ is in the $(p-3,1,1,0)$ of $SO(8)_R$
and must thus vanish for $p<3$.  So $Q^2 \O _2$ is a relevant
pseudoscalar perturbation, with $\Delta \equiv 2$, which preserves 16
supersymmetries.  This pseduoscalar deformation is associated with
fermion masses.  

We now turn to a holographic duality motivation for considering 3d RG
flows with 16 preserved supercharges.  It is expected that $M$ theory
has exact vacua with metric \review\
\eqn\mtmetric{\eqalign{ds^2&=H^{-2/3}dx_\mu ^2+H^{1/3}(dr^2+r^2d\Omega
_7^2);\cr H&=h+{2^5\pi ^2Nl_p^6\over r^6},}} where $x_\mu$ span $\IR
^{1,2}$, and there is $G_4$ field given by $G_4\sim l_p^{-3}(dx_\mu
)^{\wedge 3}\wedge dH^{-1}$, with $N$ units of $M2$ brane $G$ flux.
$M$ theory has no dilaton, which corresponds to the fact that the dual
\malda\ 3d theory has no exactly marginal operator: the theory is
conformally invariant for a fixed value of the gauge coupling constant
$g_{YM}$.  The parameter $h$ in
\mtmetric\ is again arbitrary.  Taking $h=0$, the vacuum \mtmetric\ is
exactly $AdS_4\times S^7$.  Taking $h=1$ gives the $M2$ brane metric,
which asymptotes to $\IR ^{1,10}$ for $r\rightarrow \infty$ (far from
the brane) and to $AdS_4\times S^7$ for $r\rightarrow 0$ (near horizon
limit).

In \malda\ the vacuum was originally taken to be the $h=1$ case of
\mtmetric, but then the near horizon limit, $r\rightarrow 0$, was
taken, leading to $AdS_4\times S^7$ in the limit.  One could instead
take $h=0$ from the outset.
The $h=0$ case exactly preserves $32$ supersymmetries, which combine
with the bosonic generators to give the 3d superconformal group
$SO(3,2|8)$.  This theory is expected to be exactly dual to the 3d
$SU(N)$ Yang-Mills theory with 16 supersymmetries at the RG fixed
point value of the coupling constant, $g_{YM}^*$, where the theory is
conformally invariant (and thus has an additional 16 superconformal
symmetries).  For $h\neq 0$, the vacua \mtmetric\ generally preserve
16 supersymmetries, with an additional 16 supersymmetries emerging in
the $r\rightarrow 0$ and $r\rightarrow \infty$ limits.  This case is
conjectured to be holographically dual to a field theory with a
non-trivial RG flow along which 16 supersymmetries are preserved.

$M$ theory in the vacuum \mtmetric\ is expected to be exactly
invariant under
\eqn\mtRGflow{x_{\mu}\rightarrow \lambda x_{\mu }, \quad
r\rightarrow \lambda ^{-1/2}r, \quad h\rightarrow \lambda ^{3}h,
\quad d\Omega_7\rightarrow d\Omega _7.}
This operation preserves the metric \mtmetric, $G_4\sim
l_p^{-3}(dx_\mu )^{\wedge 3}\wedge dH^{-1}$, and the other
supergravity fields.  For $h=0$, \mtRGflow\ is a symmetry which
corresponds to the dilatation generator of the superconformal group
$SO(3,2|8)$; this must be an exact gauge symmetry of the $M$ theory
vacuum in order for the theory to be holographically dual to the
exactly conformally invariant 3d field theory.  For $h\neq 0$ this
symmetry is broken, but can be regarded as being restored provided
that $h$ transforms as in \mtRGflow.  Thus, at length scale $x$, $h$
enters only via $h_{eff}\sim h/x^3$ (here $x^3\equiv (x_\mu x^\mu
)^{3/2}$), which is invariant under \mtRGflow.

We thus find that the operator by which the theory is deformed for
$h\neq 0$ has dimension $\Delta =6$ (dimension 3 in the action).  As
in the 4d case, this is twice the dimension of a marginal operator.
This was also noted in \GHKK\ via an absorption calculation.

Now consider the transformation 
\eqn\hltscale{x^\mu \rightarrow x^\mu, \quad
l_p\rightarrow \lambda l _p, \quad h\rightarrow \lambda ^{-3}h, \quad
r\rightarrow \lambda ^{3/2}r,} under which the metric \mtmetric\
transforms as $ds^2\rightarrow \lambda ^2 ds ^2$ and $G_4\sim l
_p^{-3} (dx_\mu )^{\wedge 3}\wedge dH^{-1}$ is invariant.  Because all
lengths in $M$ theory are measured relative to $l_p$, $M$ theory
should be invariant under the combined rescaling \hltscale\ of $\l _p$
and $ds ^2$.  The upshot is that $h$ should only enter via $h_{eff}=h
l_p ^3/x^3$, which is dimensionless and invariant under \mtRGflow\ and
\hltscale.

We thus propose that $M$ theory in the vacuum \mtmetric\ is dual to 
the 3d ${\cal N}=8$ field theory with Lagrangian 
\eqn\Ldefspect{{\cal L}={\cal L}_0+(const.)\ hl_p^3 \O _H,}
where ${\cal L}_0$ is the superconformal Lagrangian at $g_{YM}^*$ and
$\O _H=Q^8\O _4$, much as in the 4d case \Ldefspec.  (Again, as in 4d,
the operator $\O _4$ can actually be a combination of a single and a
double trace operator with the same quantum numbers.)
The theory \Ldefspec\  properly
preserves 16 supersymmetries and the $SO(8)_R$ symmetry, as does
\mtmetric. Also, $h$ properly couples to an operator of dimension
$\Delta =6$.  (In $D$ spacetime dimensions a scalar $\phi$ has
canonical dimension $\half (D-2)$, so $Q^8 (\phi )^4$ always has
dimension $2D$.)  The fact that $Q^4 \O _4$ continues to have
dimension $\Delta =6$, even in the deformed theory \Ldefspect\ with
$h\neq 0$, is compatible with our argument above that the dimensions
of short representation operators are not renormalized as long as 16
supersymmetries are preserved. There are also many long operators
which preserve the 16 supersymmetries and are $SO(8)_R$ singlets, but
deforming by these would not be compatible with the deformation
depending only on $h_{eff}=hl_p^3/x^3$, so we do not expect them in
\Ldefspect.

As in the 4d case, correlation functions in the proposed dual field
theory for $h\neq 0$ are given by 
\eqn\cdefii{\ev{\prod _i \O _i (x_i)}_h=\ev{e^{-(const.)\ hl_p^3
\int d^3 y\O
_H (y)}\prod _i \O _i (x_i)}_{h=0}.} The transformation \mtRGflow\
leads to the Ward identities for correlation functions
\eqn\mtRGward{\sum _{i=1}^n (x_i{\partial \over \partial x_i}+\Delta
_i)\ev{\prod _{i=1}^n \O _i(x_i)}_h=-3h{\partial \over \partial
h}\ev{\prod _{i=1}^n \O _i(x_i)}_h=\int d^3 y\ev{T^\mu _\mu (y)\prod
_{i=1}^n \O _i (x_i)}_h,}
where $\Delta _i$ are the dimensions of the operators in the perturbed
theory.  For operators in short representations, as argued above,
these dimensions are not renormalized and are independent of $h$ and
the RG scale.  In particular, the two-point function of the short
operator $\O _p$ in the perturbed theory must be of the form 
\eqn\mttpf{\ev{\O _p (x)\O _p (0)}={f_p({hl_p^3\over x^3})\over
(x^2)^{p\over 2}}.}

It would be nice to compare \cdefii\ with a detailed
partial wave absorption analysis along the lines of \GHKK.

\newsec{Six dimensional theories with 32 supercharges and RG flows
preserving 16}

Much work points to the existence of interacting 6d superconformal
field theories with 32 supercharges residing in the superconformal
group $SO(6,2|4)$.  These supercharges are the 16 of ${\cal N}=(2,0)$
supersymmetry in 6d, along with 16 superconformal partners. The
superconformal group again has short and long representations.  The
short representations of the 6d superconformal group $SO(6,2|4)$ were
obtained in \oscvid\ and are given in appendix B for convenience.  For
example, the $p=1$ multiplet is the (free) 6d ${\cal N}=(2,0)$ matter
multiplet, consisting of scalars $\O _1$ in the ${\bf 5}=(0,1)$ of
$Sp(2)_R\cong SO(5)_R$, fermions $Q\O _1$ in the ${\bf 4}=(1,0)$ of
$Sp(2)$, and self-dual tensor fields $Q^2 \O _2$ in the ${\bf 1}$ of
$Sp(2)_R$. The notation $\sqrt{a_{\alpha \beta \gamma}}$, listed in
the table for the Lorentz spin of $Q^2\O _2$, is to indicate that
these are two-form gauge fields with self-dual field strength.

The RG fixed point theory can be deformed as in \Ldef, preserving the
16 supercharges.  The $p=2$ short multiplet again contains the
currents: the $SO(5)_R$ currents are the $\Delta =5$ Lorentz vectors
$Q^2 \O _2$, the supercharges are the $\Delta = 5.5$ ``gravitinos''
$Q^3 \O_2$, and the stress tensor is the $\Delta =6$ Lorentz
``graviton'' $Q^4 \O _2$.  Again, the stress tensor remains conserved
when the theory is deformed as in \Ldef, and thus its operator
dimension is not renormalized.  The chiral ring structure of additive
anomalous dimensions for chiral superfields then ensures that the
dimensions of all operator in short representations are independent of
the RG scale, and not renormalized, in RG flows which preserve 16
supersymmetries.

Such a 6d RG flow, preserving 16 supersymmetries, can be the
holographic dual of $M$ theory vacua containing $M5$ branes.
It is expected that $M$ theory has exact vacua \review\
\eqn\mfmetric{\eqalign{ds^2&=H^{-1/3}dx_\mu ^2+H^{2/3}(dr^2+r^2d\Omega
_4^2);\cr H&=h+{\pi Nl_p^3\over r^3},}} where $x_\mu$ span $\IR
^{1,5}$, and there is $G_4$ field given by $*G_4\sim l_p^{-6}(dx_\mu
)^{\wedge 6}\wedge dH^{-1}$, with $N$ units of M5 $G$ flux.  The value
of the real parameter $h$ is again arbitrary.  For $h=1$, \mfmetric\
asymptotes to $\IR ^{1,10}$ for $r\rightarrow \infty$ and to
$AdS_7\times S^4$ for $r\rightarrow 0$.  For $h=0$, \mfmetric\ is
identically $AdS_7\times S^4$, with 32 conserved supercharges, for all
$r$.  The $h=0$ case is expected to be exactly holographically dual to
the 6d, $A_{N-1}$ type, ${\cal N}=(2,0)$ conformal field theory.  The
$h=1$ case is expected to be holographically dual to a RG flow which
preserves 16 supercharges and flows in the IR limit, $r\rightarrow 0$,
to the same CFT as in the $h=0$ case.

$M$ theory in the vacuum \mfmetric\ is expected to be exactly gauge
invariant under
\eqn\mfRGflow{x_\mu \rightarrow \lambda x_\mu, \quad r\rightarrow
\lambda ^{-2}r, \quad h\rightarrow \lambda ^6h, \quad d\Omega
_4\rightarrow d\Omega _4.}  This operation preserves the metric
\mfmetric, $*G_4\sim l_p^{-6}
(dx_\mu )^{\wedge 6}\wedge dH^{-1}$, and the other supergravity (and
$M$ theory) fields.  We again interpret this symmetry as that
associated with the dilatation generator of the superconformal group,
which is spontaneously broken for $h\neq 0$ but can be regarded as
being restored provided $h$ transforms as in
\mfRGflow.  The effective parameter which is invariant under
\mfRGflow\ is $h_{eff}\sim h/x^6$; this power of $x$ reveals that $h$
multiplies a term of exact dimension 12 in the Lagrangian density
(dimension 6 in the action).

Now consider the transformation 
\eqn\hlfscale{x^\mu \rightarrow x^\mu, \quad
l_p\rightarrow \lambda l _p, \quad h\rightarrow \lambda ^{-6}h, \quad
r\rightarrow \lambda ^{3}r,} under which the metric \mtmetric\
transforms as $ds^2\rightarrow \lambda ^2 ds ^2$ and $*G_4\sim
l_p^{-6}(dx_\mu )^{\wedge 6}\wedge dH^{-1}$ is invariant.  Because all
lengths in $M$ theory are measured relative to $l_p$, $M$ theory
should be invariant under the combined rescaling \hltscale\ of $\l _p$
and $ds ^2$.  The upshot is that $h$ should only enter via $h_{eff}=h
l_p ^6/x^6$, which is dimensionless and invariant under \mfRGflow\ and
\hlfscale.

$M$ theory with vacuum \mfmetric\ is thus expected to be
holographically dual to the 6d field theory with Lagrangian 
\eqn\Ldefspecf{{\cal L}={\cal L}_0+(const.)\ hl_p^6\O _H,}
where, ${\cal L}_0$ represents the conformal theory and again, $\O _H=
Q^8 \O _4$, which is an $SO(5)_R$ singlet and now has $\Delta = 12$,
as required above.  (\Ldefspecf\ is perhaps schematic as the theory
contains non-Abelian, self-dual tensor fields which do not have a
known, standard Lagrangian formulation.) The fact that $\O _H$ has
dimension $\Delta = 12$ even in the theory with $h\neq 0$ is
compatible with the above argument that the dimensions of operators in
short representations are not renormalized as long as the 16
supersymmetries are preserved.  The deformation of the fixed point by
a dimension $\Delta =12$ operator is compatible with the scattering
results of
\GHKK.

Corresponding to \mfRGflow, we have the Ward identities
\eqn\mfRGward{\sum _{i=1}^n (x_i{\partial \over \partial x_i}+\Delta
_i)\ev{\prod _{i=1}^n \O _i(x_i)}_h=-6h{\partial \over \partial
h}\ev{\prod _{i=1}^n \O _i(x_i)}_h=\int d^6 y\ev{T^\mu _\mu (y)\prod
_{i=1}^n \O _i (x_i)}_h,}
where $\Delta _i$ are the dimensions of the operators in the perturbed
theory.  For operators in short representations, as argued above,
these dimensions are not renormalized and are independent of $h$ and
the RG scale.  In particular, the two-point function of the short
operator $\O _p$ in the perturbed theory must be of the form
\eqn\mttpf{\ev{\O _p (x)\O _p (0)}={f_p({hl_p^6\over x^6})\over
(x^2)^{2p}}.}

\newsec{Non-commutative geometry and maximally supersymmetric RG flows}

There has been recent interest in gauge theories in non-commutative
spaces; such theories arise in the world-volume of branes with
background $B$ field (or, in $M$ theory, the $C$ field).  The
non-commutativity of space-time, $[x^\mu , x^\nu ]=i\theta ^{\mu
\nu}$, introduces a length scale via the parameter $\theta ^{\mu
\nu}$, which clearly has mass dimension $\Delta =-2$.  Thus the maximal
supersymmetry for $\theta ^{\mu \nu}\neq 0$ is the 16 ordinary
supercharges, without the superconformal symmetries, and there is a RG
flow to the IR, where $\theta ^{\mu
\nu}$ becomes irrelevant.  The IR fixed points are the ordinary,
superconformal ${\cal N}=4$ Yang-Mills theories on commutative space,
with 32 supercharges.  The non-commutativity thus leads to maximally
supersymmetric RG flows, of precisely the type discussed in the
previous sections.

\nref\swnc{N. Seiberg and E. Witten, hep-th/9908142.}
Indeed, it was argued in \swnc\ that non-commutative gauge theories
can be related to ordinary gauge theories by a field-redefinition of
the gauge field strength, order-by-order in the parameter $\theta
^{\mu \nu}$; the explicit change of variables can be found in
sect. 3.1 of \swnc. Thus the non-commutative theory with 16
supercharges is equivalent to an ordinary theory with 16 supercharges
with higher dimension terms in the Lagrangian, exactly as in \Ldef,
coming from the field redefinition; these terms are weighted by powers
of $\theta ^{\mu \nu}$ and become irrelevant in the IR.

Suppose e.g. that we start with a 4d (other $d$ are similar)
Lagrangian which is formally the same as the ${\cal N}=4$
superconformal Lagrangian, but is not conformally invariant simply
because the space-time is non-commutative.  Via the change of
variables of \swnc, this should be equivalent to the ordinary gauge
theory with Lagrangian
\eqn\Ldefnci{{\cal L}={\cal L}_0+\theta ^{\alpha \beta}[Q^2\overline 
Q^4 {\cal O}_3]_{\alpha \beta}+ \theta ^{\dot \alpha \dot \beta}
[\overline Q ^2 Q^4{\cal 
O}_3]_{\dot \alpha \dot \beta}+ (const.)(\theta ^{\mu \nu})^2
\O _H +\dots,} 
with $\theta ^{\alpha \beta}$ and $\theta ^{\dot \alpha \dot \beta}$
the Lorentz spin $(1,0)$ and $(0,1)$, respectively, parts of $\theta
^{\mu \nu}$.  The operators $[Q^2\overline Q^4 {\cal O}_3]_{\alpha
\beta} \sim (\Tr F^3)_{\alpha \beta}+\dots$ and $[\overline Q ^2
Q^4{\cal O}_3]_{\dot \alpha \dot \beta}$ are dimension $\Delta =6$
short operators and $\O _H=Q^4\overline Q^4 \O _4$ is the $\Delta =8$
short operator; all of these operators are $SU(4)_R$ flavor singlets
and annihilated by the 16 supersymmetries, so
\Ldefnci\ respects the expected symmetries of non-commutative ${\cal N}=4$.
In principle, there should be terms in \Ldefnci\ at higher orders of
$\theta ^{\mu \nu}$.  However, any terms of higher order in $\theta$
in \Ldefnci\ must be long operators which preserve the 16
supersymmetries and $SU(4)_R$, as there are no other short operators
which respect these symmetries.
The fact that the dimensions of the short operators appearing in 
\Ldefnci\ are not renormalized along the
RG flow of \Ldefnci\ is consistent with the expected non-renormalization
of the dimension, $\Delta =-2$, of the parameter $\theta ^{\mu
\nu}$ appearing in $[x^\mu, x^\nu]$.  
On the other hand, the coefficients of possible long operators
appearing in
\Ldefnci\ would have to have quantum-corrected anomalous dimensions, to
compensate for the anomalous dimensions of the long operators to which
they couple.  

The arguments of the previous sections apply directly here.  The 16
supersymmetry RG flows associated with non-commutative geometry will
have short operators, with non-renormalized dimensions along the
entire RG flow.  In the 4d case,  the gauge coupling must thus remain
constant along the RG flow, exactly as in fig. 1.

\nref\HI{A. Hashimoto and N. Itzhaki, hep-th/9907166.}
\nref\MR{J.M. Maldacena and J. Russo, hep-th/9908134.}
\nref\AOS{M. Alishahiha, Y. Oz, and M.M. Sheikh-Jabbari,
hep-th/9909215.}  String (or M theory) vacua with non-zero $B$ (or
$C$) field and their conjectured holographic duality to world-volume
field theories were discussed in \refs{\HI - \AOS}.  These vacua have
16 supersymmetries.  For example, the case associated with IIB vacua
with D3 brane charge (the M2 and M5 cases are similar) has \refs{\HI,
\MR}, in our earlier notation, exact vacua:
\eqn\noncv{\eqalign{
ds^2_{str}&=H^{-1/2}[f_1(dx_0^2+dx_1^2)+f_2(dx_2^2+dx_3^2)]
+H^{1/2}d\vec y\cdot d\vec y,\cr H&=h+{4\pi g_0\alpha ^{\prime 2}\over
\cos \theta _1
\cos \theta _2}\sum _{i=1}^k {N_i\over |\vec y - \vec a_i|^4}, 
\quad f_{j=1,2}^{-1}=\sin ^2 \theta _j H^{-1}+\cos ^2 \theta _j,\cr
e^\phi &= g_0\sqrt{f_1f_2}, \qquad 2\pi
\alpha 'B_{01}=\tan \theta _1 H^{-1}f_1, \quad 2\pi \alpha
'B_{23}=\tan \theta _2 H^{-1}f_2.}}  The $\theta _j$ are dimensionless
free parameters, as is $g_0\geq 0$.  As in the previous sections,
we conjecture that IIB string theory in the general vacuum
\noncv\ is holographically dual to a deformed
${\cal N}=4$ supersymmetric theory
\Ldef, with 16 supersymmetries, even away from the
near-horizon limit of \refs{\HI-\AOS}.  

The $\vec y$ dependent dilaton in \noncv\ seems to contradict our
arguments of sect. 2 that $g_{YM}$ must be independent of the RG
scale.  Fortunately, it is wrong 
\foot{I am grateful to N. Seiberg for pointing this out to me.}
here to simply identify $e^\phi$ with $g_{YM}^2$.  (Rather, $e^\phi$ 
is the suppression factor for non-planar diagrams \MVS.)

The point is that the supergravity solution \noncv\ should be
regarded as giving the {\it closed string} quantities, whereas 
the worldvolume gauge theory is sensitive to {\it open string}
quantities \swnc. We'll assume that we can directly apply the
formulae of \swnc, 
\eqn\swform{\eqalign{G_{ij}&=g_{ij}-(2\pi \alpha ')^2(Bg^{-1}B)_{ij},\cr
\theta ^{ij}&=2\pi \alpha '\left({1\over g+2\pi \alpha 'B}\right)^{ij}_A\cr
g_{YM}^{-2}&=e^{-\phi}\left({\det (g+2\pi \alpha ' B)\over
\det (g-(2\pi \alpha ')^2Bg^{-1}B)}\right)^{1/2},}}
to the non-flat background \noncv.  Doing so, we obtain for the
open string metric and $\theta ^{ij}$,
\eqn\opennc{\eqalign{dS^2_{open}&=H^{-1/2}[\sec ^2\theta _1(dx_0^2+
dx_1^2)+\sec ^2\theta _2(dx_2^2+dx_3^2)]+H^{1/2}d\vec y\cdot d\vec y,\cr
\theta ^{01}&=-\pi \alpha ' \sin 2\theta _1\qquad \theta ^{23}=-\pi \alpha '
\sin 2\theta _2.}}
The worldvolume gauge coupling obtained via \swform\ is
\eqn\gymrr{g_{YM}^{-2}={1\over g_0\sqrt{f_1f_2}}\sqrt{\prod
_{i=1,2}{f_i^2 (H^{-1}+H^{-2}\tan ^2 \theta _i)\over (H^{-1/2}f_i+\tan
^2 \theta _i H^{-3/2}f_i)^2}}={\cos \theta _1 \cos \theta _2 \over
g_0}.}

Remarkably, the non-trivial functions $f_i$ of the AdS bulk coordinate
$\vec y$ appearing in \noncv\ all completely cancel out of the open
string quantities!  Up to constant rescalings, $dS_{open}^2$ and
$g_{YM}^{-2}$ are unaffected by the $B$ field.  The $\vec y$
independence of the $\theta ^{ij}$ in
\opennc\ shows that these field theory parameters are $\Delta =-2$
constants, which are otherwise not renormalized.  This is as should
have been expected from $[x^\mu ,x^\nu]=i\theta ^{\mu \nu}$.  The
$\vec y$ independence of $g_{YM}^{-2}$ in \gymrr\ is in agreement with
the general considerations of sect. 2: the dual field theory RG flow
is indeed as in fig. 1, with $g_{YM}$ constant along the entire RG
flow.

\bigskip
\centerline{{\bf Acknowledgments}}

I would like to thank P. Fendley, M.R. Plesser, W. Skiba,
M.J. Strassler, and W. Taylor for discussions. This work was partly
supported by UCSD grant DOE-FG03-97ER40546.  This work was initiated
while I was on leave {}from UCSD and visiting the IAS, where I was
fully supported by an IAS grant {}from the W.M. Keck Foundation; I
would like to thank the IAS for this support and for much hospitality.

\appendix{A}{The short multiplets of the 3d
${\cal N}=8$ superconformal group (see \osciiid)}
$$\matrix{{\rm form} &&{\rm spin/parity}&&\Delta && SO(8)_R\cr 
\O _p&&0_+&&\half p&&(p,0,0,0)\cr
Q \O _p&&\half &&\half p+\half &&(p-1,0,1,0)\cr
Q^2\O _p&&1_-&&\half p+1 &&(p-2,1,0,0)\cr
Q^3\O _p&& \f{3}{2}&&\half p +\f{3}{2}&&(p-2,0,0,1)\cr
Q^4\O _p&&2&&\half p+2&&(p-2,0,0,0)\cr
Q^2\O _p&& 0_-&&\half p+1&&(p-2,0,2,0)\cr
Q^3\O _p&&\half &&\half p+\f{3}{2}&&(p-3,1,1,0)\cr
Q^4\O _p&&1_+&&\half p+2&&(p-3,0,1,1)\cr
Q^5\O _p&&\f{3}{2}&&\half p+\f{5}{2}&&(p-3,0,1,0)\cr
Q^5\O _p&&\f{1}{2}&&\half p+\f{5}{2}&&(p-4,1,0,1)\cr
Q^7\O _p&&\half&&\half p+\f{7}{2}&&(p-4,0,0,1)\cr
Q^6\O _p&&1_-&&\half p+3&&(p-4,1,0,0)\cr
Q^4\O _p&&0_+&&\half p+2&&(p-4,0,0,2)\cr
Q^6\O _p&& 0_-&&\half p+3&&(p-4,0,0,2)\cr
Q^8\O _p&&0_+&&\half p +4&&(p-4,0,0,0)\cr}$$

\appendix{B}{The short multiplets of the 6d
${\cal N}=(2,0)$ superconformal group (see \oscvid)}
$$\matrix{{\rm form} &&{\rm spin}&&\Delta && Sp(2)_R\cr 
\O _p&&{\rm scalar}&&2p&&(0,p)\cr
Q\O _p&&{\rm spinor}&&2p+\half&&(1,p-1)\cr
Q^2\O _p&&\sqrt{a_{\alpha \beta \gamma}}&&2p+1&&(0,p-1)\cr
Q^2\O _p&&\rm{vector}&&2p+1&&(2,p-2)\cr
Q^3\O _p&&\rm{gravitino}&&2p+\f{3}{2}&&(1,p-2)\cr
Q^4\O _p&&\rm{graviton}&&2p+2&&(0,p-2)\cr
Q^3\O _p&&\rm{spinor}&&2p+\f{3}{2}&&(3,p-3)\cr
Q^4\O _p&&a_{\alpha \beta}&&2p+2&&(2,p-3)\cr
Q^5\O _p&&\rm{gravitino}&&2p+\f{5}{2}&&(1,p-3)\cr
Q^6\O _p&&\sqrt{a_{\alpha \beta \gamma}}&&2p+3&&(0,p-3)\cr
Q^4\O _p&&\rm{scalar}&&2p+3&&(4,p-4)\cr
Q^5\O _p&&\rm{spinor}&&2p+\f{5}{2}&&(3,p-4)\cr
Q^6\O _p&&\rm{vector}&&2p+3&&(2,p-4)\cr
Q^7\O _p&&\rm{spinor}&&2p+\f{7}{2}&&(1,p-4)\cr
Q^8\O _p&&\rm{scalar}&&2p+4&&(0,p-4).\cr}$$

\listrefs \end